\documentclass[%
 reprint,
 amsmath,amssymb,
 aps,
]{revtex4-2}

\usepackage{hyperref}
\usepackage{graphicx}
\usepackage{dcolumn}
\usepackage{bm}
\usepackage[usenames, dvipsnames]{color}
\usepackage{color}
\usepackage{amsmath}

\DeclareMathOperator*{\argmin}{arg\,min}

\begin{document}

\title{Landau Theory for the Mpemba Effect Through Phase Transitions}

\author{Roi Holtzman}
\affiliation{Department of Physics of Complex Systems, Weizmann Institute of Science, Rehovot, 76100, Israel}
\author{Oren Raz}
 \email{oren.raz@weizmann.ac.il}
\affiliation{Department of Physics of Complex Systems, Weizmann Institute of Science, Rehovot, 76100, Israel}

\date{\today}

\begin{abstract}

The Mpemba effect describes the situation in which a hot system cools faster than an identical copy that is initiated at a colder temperature. In many of the experimental observations of the effect, e.g. in water and clathrate hydrates, it is defined by the phase transition timing. However, none of the theoretical investigations so far considered the timing of the phase transition, and most of the abstract models used to explore the Mpemba effect do not have a phase transition. We use the phenomenological Landau theory for phase transitions to identify the second order phase transition time, and demonstrate with a concrete example that a Mpemba effect can exist in such models.

\end{abstract}

\maketitle

\section{Introduction}
\label{sec:introduction}
Under appropriate conditions, a cup of hot water may freeze faster than an identical cup of cold water. This counter intuitive phenomenon was documented as early as 2300 years ago \cite{Aristotele350,Jeng2006}, but is named after A. Mpemba -- a Tanzanian school student that rediscovered it in the 60's \cite{Mpemba1969}. Several mechanisms were suggested to explain the Mpemba effect in water, including: evaporation \cite{Theory_evaporation_kell1968freezing,Mirabedin2017}, dissolved gases and solids \cite{Disolved_katz2009hot}, convection flow \cite{Vynnycky2015}, super-cooling \cite{Auerbach1998} and anomalous relaxation of hydrogen bonds \cite{Zhang2014}.

In recent years the term ``Mpemba effect" was extended, and it is now used to describe a wide range of non-monotonic relaxation phenomena. These include experimental observations of hot systems that undergo a phase transition before cold systems in non-water substances (Polymers \cite{Hu2018}, Clathrate hydrates \cite{paper:hydrates}), as well as in other types of phase transitions (Magnetic transition in alloys \cite{Chaddah2010} and various spin models \cite{baity2019mpemba,yang2020non,vadakkayil2021should,nava2019lindblad,teza2021relaxation}), relaxation towards equilibrium without a phase transition that is non-monotonous in the initial temperature \cite{Lu2017,Gal2020,Klich2019,Walker2021,busiello2021inducing} and similar effects in relaxation towards a non-equilibrium steady states in driven molecular gas models \cite{Lasanta2017,biswas2020mpemba,takada2021mpemba,mompo2021memory,megias2022thermal,biswas2022mpemba}.

Significant progress was recently achieved in understanding non-monotonic relaxations towards both equilibrium and non-equilibrium states, including a careful mathematical formulation of the problem \cite{Lu2017,chetrite2021metastable}, prediction of an inverse Mpemba effect where a cold system heats up faster than a hot one \cite{Lu2017,Lasanta2017}, and of the ``strong Mepmba effect" where an exponentially faster relaxation can be achieved from specific initial temperatures \cite{Klich2019}. Some of these theoretical predictions were experimentally verified in \cite{Kumar2020,kumar2022anomalous}. These results {focus} on the long time behavior of the system, and are therefore not informative for experiments and numerical simulations where the system undergoes a phase transition after a finite time, as in water, clathrate hydrates, polymers and magnetic alloys. 

In this work we present a theoretical model for the Mpemba effect through a second order phase transition. We first define, in the context of Landau theory, the exact phase of the system throughout its relaxation process, which is naturally far from equilibrium. With this definition, the phase transition can be associated with a concrete time for any initial condition. Using this ``time to phase transition", a Mpemba effect can be defined and identified. A specific mechanism for the Mpemba effect through such a transition is then demonstrated with a concrete example of a Landau free energy.  

Throughout the manuscript we limit the discussion to the Mpemba effect through a second order phase transition. To keep the description simple, from this point on we use the term \emph{Mpemba effect} to describe the scenario in which it takes less time for an initially hot system to undergo a second order phase transition in comparison to an initially colder system.

\section{Non-equilibrium phase transition at finite time}
\label{sec:non-equilibrium-phase}

The existence of the Mpemba effect considered in this manuscript is determined by the time it takes the system to undergo a second order phase transition as a function of the initial temperature, when the system is quenched to a cold environment. However, during the relaxation process the system is generically not in an equilibrium state associated with any temperature, and it is not always possible to define the phase of the system in these cases. Moreover, in many types of dynamics (e.g. coarsening dynamic \cite{bray2002theory}), the phase transition happens only in the infinite time limit. Other finite time phase transitions out of equilibrium  are not associated with the mean value of the order parameter, but rather with its fluctuations \cite{meibohm2021finite}, and are therefore not useful in the context of the Mpemba effect. For these reasons, we first suggest a simple definition for the moment in time at which the phase transition happens when the system is coupled to an infinite, memory-less heat bath, and which is finite in some relevant class of models. In the spirit of the Landau theory, we consider mean-field theories, i.e. models without any spatial dependence. 

\subsection{The Phases of the System}

Consider a system that can be characterized by a set of macroscopic parameters $x_1, \ldots, x_n$. These are often represented, for short handed, as $\vec{x} = (x_1,...,x_n)$. Some of these parameters, say $x_1, \ldots, x_m$ are the \emph{order parameters}, i.e. their value determines the phase of the system. For simplicity, in what follows we assume that there is a \emph{single} order parameter in the system, $x_1$. Upon quenching the system to a different temperature, the macroscopic parameters evolve in the configuration space of $\vec{x}$ towards their new equilibrium value. We denote the equilibrium value of $\vec x$ that corresponds to some temperature $T$ by $\vec{x}^{\rm eq}(T)$.  

Typically, an order parameter of a second order phase transition is defined such that it is zero in one phase and non-zero in another phase. In our case the two phases are therefore characterized by the equilibrium value of the order parameter,
\begin{equation}
\label{eq:x1-order-disorder}
x_1^{\rm eq}(T)
\begin{cases}
  = 0, \quad \text{disordered, } & T\geq T_c \\
  \neq 0, \quad \text{ordered, } & T< T_c
\end{cases}
\end{equation}
where $T_c$ is the critical temperature of the model. For simplicity we consider here the common case where above $T_c$ the system is disordered and consequently $x_1^{\rm eq}=0$, whereas below $T_c$ the system is in one of the ordered phases and $x^{\rm eq}_1$ is either negative or positive. We comment on the less common case where the system is ordered for $T>T_c$ and disordered for $T<T_c$ in Sec. \ref{Sec:discussion}.

We assume that both the equilibrium and non-equilibrium values of these order parameters are determined by the Landau free energy which we discuss in what follows.

\subsection{Landau Free Energy}
\label{sec:effect-free-energy}
We denote the Landau free energy of the system, which is defined for any value of the macroscopic parameters $\vec x$, by $f(\vec x; T)$. The equilibrium configuration of the system at temperature $T$ corresponds to the global minimum of $f(\vec x; T)$, namely
\begin{equation}
\label{eq:x-eq}
\vec{x}^{\rm eq}(T)  = \argmin_{\vec x} f(\vec x; T).
\end{equation}

To date, there is no single theory for the dynamics of the macroscopic parameters under all non-equilibrium conditions, but several models are often used to describe specific non-equilibrium scenarios. To describe the relaxation towards an equilibrium state, we use the common assumption \cite{bray2002theory} that the dynamic of the $x_i$ parameters is given by the negative gradient of $f(\vec{x}; T)$ and a stochastic noise,
\begin{equation}
\label{eq:langevin}
\dot{x}_i = - \frac{\partial f(\vec{x}; T)}{\partial x_{i}} + \xi_i.
\end{equation}
This form corresponds to Model A in the classification of Hohenberg and Halperin (see Eqs. (4.1) in Ref.~\cite{hohenbergTheoryDynamicCritical1977}).
$\vec{\xi}$ is a thermal noise associated with the external bath temperature, such that the equilibrium probability distribution is the expected Boltzmann distribution. Moreover, at each extremum  point of $f(\vec{x}; T)$, the first term on the right hand side of the above equation vanishes. Thus, without a noise term, all of these points were stationary. This is the desired property of the minima of the free energy, but not of its other types of extremum points. The noise term remedies this issue: the system remains in the vicinity of its minima, given that the noise is not too strong, but not near any other types of fixed points.

\subsection{The Phase of a Non-Equilibrium State}
\label{sec:phase-non-equil}

Consider a quench protocol that takes a system that is prepared in equilibrium at $T_0>T_c$, corresponding to the disordered phase, and connects it to a bath at $T_b<T_c$, corresponding to the ordered phase. Thus the system is initiated at $ x_1(t=0)  = 0$, and ends at $ x_1(t=\infty)  \neq 0$.

To identify the phase transition time, it is natural to consider the moment in time at which the system has changed from the disordered phase to the ordered phase, manifested in the growth of $|x_1|$. By Eq.~\eqref{eq:langevin}, the dynamic of $x_1$ is dictated by the effective free energy $f(\vec{x}; T)$ that acts as a potential that guides the system towards its equilibrium state. As we assume spontaneous symmetry breaking around $x_1=0$, namely that $\partial_{x_1}f(x_1=0,x_2,...,x_n; T)=0$, the growth of $|x_1|$ is determined by the second derivative in the $x_1$ direction of $f(\vec{x}; T)$ around the hyper-plane $(x_1=0, x_2, \ldots, x_n)$. For $\partial_{x_1}^2 f(x_1=0, x_2, \ldots, x_n;T) > 0$, the effective free energy confines $x_1$ around $x_1=0$, whereas for $\partial_{x_1}^2 f(x_1 = 0, x_2, \ldots, x_n;T) < 0$, the effective free energy pushes $x_1$ towards a non-zero value. In the latter case, the specific noise realization breaks the symmetry and dictates whether $x_1$ becomes positive or negative.

We therefore define the phase transition time $t_c$ as the smallest time $t$ that solves the following equation:
\begin{eqnarray} \label{Eq:PhaseTRansTime}
\partial_{x_1}^2f\Big(\langle \vec x(t)\rangle;T_b \Big) =0, 
\end{eqnarray}
where $T_b$ is the bath temperature, $\vec x(t=0)$ is sampled from the equilibrium corresponding to the initial temperature, $\vec{x}(t)$ follows the dynamic in Eq.~\eqref{eq:langevin} and where $\langle \cdots \rangle$ denotes averaging over the noise realizations sampled from the bath.

In the Supp. Info. (App. \ref{sec:appendix-phase-trans-finite-time}) we demonstrate this definition of $t_c$ for a concrete, microscopic model -- the mean field anti-ferromagnetic Ising model under the Glauber dynamics \cite{Glauber1963}.  We show the following properties of the phase transition time $t_c$: (i) It can be defined in a more general setting of relaxation dynamics than considered in Eq.~\eqref{eq:langevin}, namely a dynamic which is not the gradient of the free energy; (ii) It is finite even in the thermodynamic limit; and (iii) Its variance decreases with the system size, and therefore $t_c$ is well behaved in the thermodynamic limit.

\section{The Mpemba effect}
\subsection{Definition}
\label{Sec:MpembaDeff}

Once the exact time at which the phase transition happens has been defined, the definition of the Mpemba effect follows. We say that a Mpemba effect exists in the system if: (i) The system has a phase transition at some critical temperature $T_c$, such that Eq.~\eqref{eq:x1-order-disorder} holds; (ii) There exist two initial temperatures above the critical temperature, $T^i_{\rm hot}>T^i_{\rm cold}>T_c$, and a final temperature below it, $T^f<T_c$, such that when quenched to the cold temperature $T^f$, the time $t_c$ to reach the phase transition as defined in Eq.~\eqref{Eq:PhaseTRansTime} is smaller for the system initiated at the higher temperature $T^i_{\rm hot}$ than for a system initiated at the lower temperature $T^i_{\rm cold}$.

\subsection{Systems With One Macroscopic Parameter}
\label{sec:1d-phase-transition}

In the case of $n=1$, there is only one macroscopic parameter $x_1$ which is therefore  also  the order parameter. This means that the effective free energy has, in the close vicinity of the phase transition, the familiar form 
\begin{equation}
  \label{eq:effective-f-1d}
f(x_1 ; T) = a_0(T - T_c) x_1^2 + b_0 x_1^4 ,
\end{equation}
where $a_0, b_0 > 0$ are phenomenological constants and the free energy is expanded around the phase transition point only to the fourth order in $x_1$.
Indeed, for $T>T_c$ there is one minimum at $x_1=0$ corresponding to the disordered phase, whereas for $T<T_c$ there are two minima at $x_1=\pm \sqrt{a_0(T-T_c) / b_0}$. 

For any initial temperature $T_0>T_c$, the initial configuration of the system is given by $\left\langle x_1 (t=0) \right\rangle_{T_0}=0$.
For any bath temperature $T_b<T_c$ the effective free energy is unstable at $x_1=0$, namely $\partial_{x_1}^2 f(\langle x_1(t=0)\rangle; T_b)|_{x_1=0} < 0$, and therefore by the definition of $t_c$, Eq.~\eqref{Eq:PhaseTRansTime}, the phase transition happens instantaneously.

Hence, in mean field models that have a single macroscopic parameter, as the ferromagnetic Ising model, all hot temperatures cross the phase transition at the zero time, and consequently there cannot be a Mpemba effect as defined above.

\subsection{Systems With Two Macroscopic Parameters}
\label{sec:2d-phase-transition}

As we next show, when the configuration space has at least two dimensions, the phase transition can occur at some non-zero time. Therefore, in such systems the Mpemba effect is plausible.

Consider the case where there are two macroscopic parameters, $\vec x = (x_1, x_2)$, with $x_1$ being the order parameter that corresponds to a spontaneous symmetry breaking, whose value determines the phase of the system according to Eq.~\eqref{eq:x1-order-disorder}. We assume that $f$ is symmetric with respect to $x_1=0$ at all temperatures. In this case, for fixed values of $x_2$ and $T$, $f(x_1, x_2; T)$ as a function of $x_1$ has either a minimum or a maximum at $x_1=0$. Thus, it is possible  that for some temperature $T$ the free energy surface has a range of \(x_2\) values for which \(f(x_1, x_2; T)\) is stable with respect to \(x_1\), and a different range of \(x_2\) values for which it is unstable with respect to \(x_1\). If in this case a system is initiated in the stable region, and its dynamic, governed by Eq.~\eqref{eq:langevin}, guides the system to the unstable region, a phase transition happens in the system after a finite time.

\begin{figure*}[htbp]
\centering
\centerline{\includegraphics[width=1.0\linewidth]{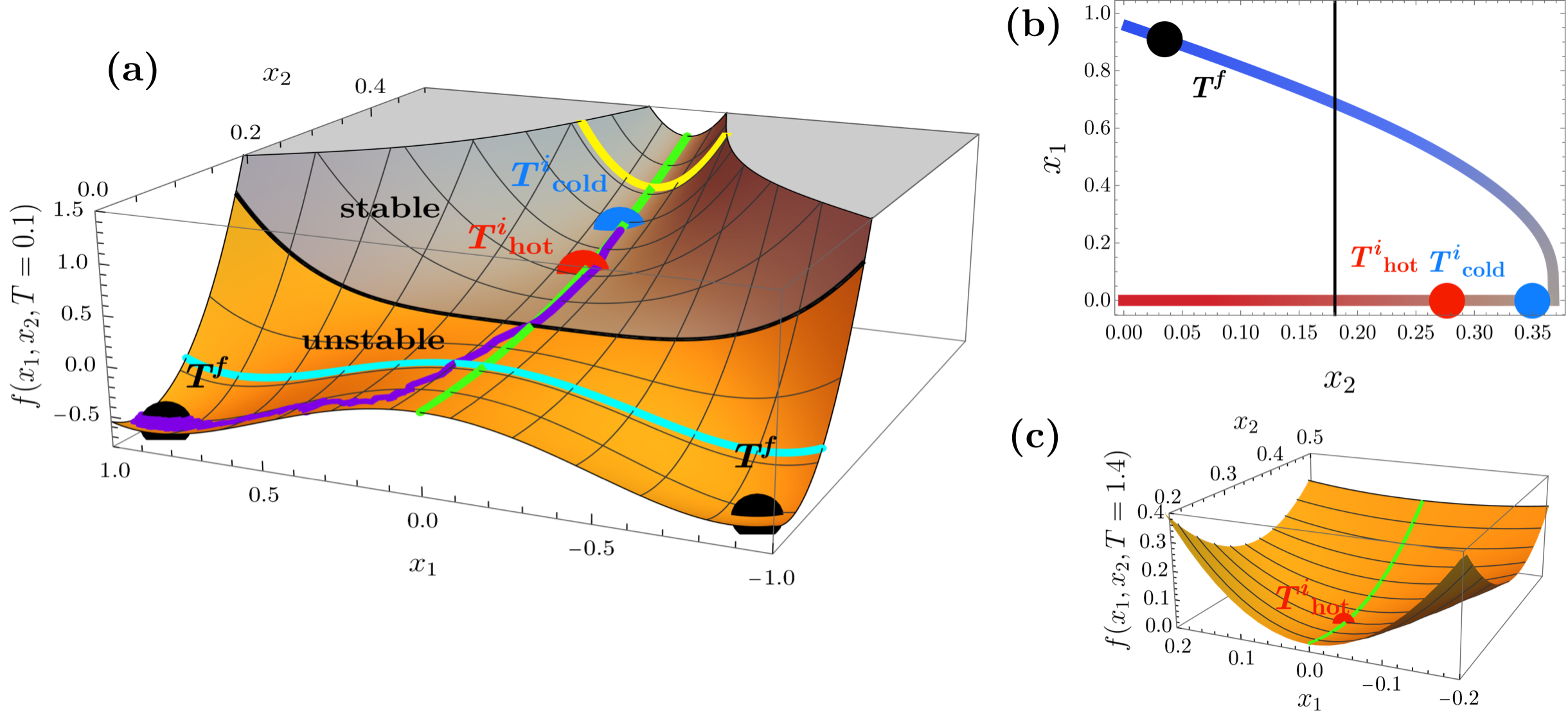}}
\caption{ A concrete example of a free energy demonstrating both a phase transition occurring after a finite time and a Mpemba effect. The explicit form of the free energy $f(x_1, x_2; T)$ is given in Eqs. (\ref{eq:f-landau-2d-example}, \ref{eq:gamma}, \ref{eq:psi2}, \ref{eq:a}). \\
(a) The free energy surface evaluated at $T=0.1<T_c$. The green line, $x_1=0$, corresponds to all equilibrium values at the disordered phase. The purple line is an example of a trajectory which starts at $\vec x^{\rm eq} (T_{\rm cold})$ (the blue dot), and follows Eq.~\eqref{eq:langevin} for some random noise realization. The stability in the $x_1$ direction along the $x_1=0$ line changes throughout the trajectory -- the black line separates between the stable region (gray) and the unstable region (orange). Crossing this line defines the phase transition, and it happens at a non-zero finite time. The red and blue dots correspond to hot and cold initial equilibria in the disordered phase, i.e. to $\vec x^{\rm eq}(T^i_{\rm hot})$ and $\vec x^{\rm eq}(T^i_{\rm cold})$, respectively. The black dots correspond to the symmetric equilibria of the final energy $T^f = 0.1$.
Observing the trajectory governed by the free energy surface, it is clear that as the colder initial condition must first reach the hotter initial condition, the colder initial condition takes more time, and so this system exhibits the Mpemba effect.\\
(b) The equilibrium line $\vec{x}^{\rm eq}(T)$ of the free energy  determined by Eq.~\eqref{eq:x-eq}. Temperatures range from $T=0$ (blue) to $T=\infty$ (red). The black, blue and red dots correspond the same dots of panel (a), i.e. to $\vec x^{\rm eq}(T^f), \vec x^{\rm eq}(T^i_{\rm cold})$ and $\vec x^{\rm eq}(T^i_{\rm hot})$, respectively. It is clear that the hotter initial condition is closer to $T^f$ than the colder initial condition. The black line corresponds to the black line in panel (a), namely it separates the stable region and the unstable regions at $T^f=0.1$. \\
(c) The free energy surface evaluated at $T=1.4>T_c$. This temperature is denoted as $T^i_{\rm hot}$ and therefore the minimum of $f(\vec x; T^i_{\rm hot})$ is obtained by $\vec x^{\rm eq}(T^i_{\rm hot})$ as seen in the plot. The green line corresponds to the symmetry line $x_1=0$. It is demonstrated that $x_1=0$ is stable for all values of $x_2$ at this temperature. The stability of $x_1=0$ holds for all temperatures above $T_c$.}
\label{fig:free-energy}
\end{figure*}

The above  scenario is demonstrated by the free energy surface plotted in Fig.~\ref{fig:free-energy}(a). This specific free energy is constructed in Sec.~\ref{sec:example-mpemba2-2d}.
The black line, which we denote by $x_2 = x_2^{\rm st}$, separates between  a stable region (gray) and an unstable region (orange). 
Initial conditions in the stable region, such as the blue and red dots, would stay confined around the $x_1=0$ line, for most noise realizations.
Once the system crosses to the unstable region, $x_1$ is no longer confined, and the noise pushes the system towards one of the minima.
An example for such a trajectory is plotted in purple in Fig.~\ref{fig:free-energy}(a).
Therefore, in this two-dimensional configuration space different initial conditions cross the phase transition at different non-zero finite times. This implies that a Mpemba effect is plausible. In the next section we provide a concrete example that demonstrates the Mpemba effect as defined above.

\section{Example of a System with a Mpemba Effect}
\label{sec:example-mpemba2-2d}

\subsection{Required Features for the Mpemba Effect}

In this section we show that a Mpemba effect can exist in a system with two macroscopic  parameters, namely for $\vec{x}=(x_1, x_2)$. Before providing a concrete example, let us first explain the basic idea, demonstrated in Fig.~\ref{fig:free-energy}. To this end, we consider for each temperature $T$ the corresponding 2d free energy surface,  $f(x_1,x_2;T)$. The global minima of this surface dictates the equilibrium values, $x_1^{\rm eq}(T)$ and $x_2^{\rm eq}(T)$. At some critical temperature $T_c$, $x_1^{\rm eq}(T)$ changes from $x_1^{\rm eq}=0$ at $T>T_c$ to $x_1^{\rm eq} \neq 0$ at $T<T_c$. We denote $x_2^* \equiv x_2^{\rm eq}(T_c)$, namely the value of $x_2^{\rm eq}$ at the critical temperature. 

To observe the Mpemba effect, the system is initiated at two different temperatures in the hot temperature phase, where $x_1^{\rm eq}=0$. The difference between the initial conditions at $T^i_{\rm hot}$ and $T^i_{\rm cold}$ is therefore not in their $x_1$, but rather in their $x_2$ values. We construct $f(\vec x;T)$ such that $x_2^{\rm eq}(T)$ is non-monotonic in $T$, and is maximal at $T_c$, so that $x_2^*$ is the maximal equilibrium value of $x_2$. This feature exists for example in the mean field anti-ferromagnet Ising model in the presence of a weak external magnetic field (see \cite{vivesUnifiedMeanfieldStudy1997} as well as the Supp. Info.~\ref{sec:appendix-phase-trans-finite-time}). The ``equilibrium line" of the model described in what follows is plotted over the $(x_1,x_2)$ plane in Fig.~\ref{fig:free-energy}(b), and it demonstrates this feature.
The non-monotonicity of $x_2^{\rm eq}(T)$ gives the following property. Consider the black, blue and red dots in Fig.~\ref{fig:free-energy}(b). The three dots correspond to the equilibria of  $T^f< T^i_{\rm cold}< T^i_{\rm hot}$, respectively.
It is seen that between the two points in the disordered phase, $x_2^{\rm eq}(T^i_{\rm hot})$ and $x_2^{\rm eq}(T^i_{\rm cold})$, the one that is closer to the final equilibrium state $x_2^{\rm eq}(T^f)$ is actually the hotter point $x_2^{\rm eq}(T^i_{\rm hot})$. This is a crucial feature for the Mpemba effect in this system.

The next feature we describe regards the stability of the system, determined by $f(x_1, x_2; T)$, with respect to its symmetric coordinate $x_1$. At the final cold temperature $T^f$, the free energy surface should have the following property: the symmetry line of the system, $x_1=0$ (the green line in Fig.~\ref{fig:free-energy}(a)), has two different regions --- one stable in the $x_1$ direction and the other unstable in the $x_1$ direction. 
The stability of the system with respect to $x_1$ on the symmetry line $x_1=0$ is given by the sign of the second derivative $\partial^2_{x_1} f|_{x_1=0}$: if it is positive the system is stable, and if it is negative the system is unstable.
Having two different regions of stability, means that the stability changes as a function of $x_2$.
The simplest setting for these two regions is having one point, which we denote the stability point $x_2^{\rm st}$, separating between the stable and unstable regions.
The presented example adheres to this simple setting, as shown in Fig.~\ref{fig:free-energy}(a) --- the black line is located at $x_2=x_2^{\rm st}$ and it separates between the stable region (gray) and the unstable region (orange).
These conditions are summarized by
\begin{eqnarray}
\label{eq:f-stability-cond}
\partial^2_{x_1} f (x_1 =0, x_2; T) 
\begin{cases}
  <0, & \hbox{for } 0 < x_2 < x_2^{\rm st} \\
  =0, & \hbox{for } x_2 = x_2^{\rm st} \\
  >0, & \hbox{for } x_2^{\rm st} < x_2 < x_2^{*}.
\end{cases}
\end{eqnarray}

\subsection{The Form of the Landau Free Energy}

Let us demonstrate the above idea with a concrete construction of $f(x_1,x_2;T)$. The Landau free energy has the following form:
\begin{eqnarray}
\label{eq:f-landau-2d-example}
f(x_1,x_2; T) &=& \gamma(x_2, T) + \psi(x_2, T) x_1^2 +  x_1^4.
\end{eqnarray}
It is composed of $x_1^0$, $x_1^2$ and  $x_1^4$ terms, which is the simplest form for a system with a second order phase transition for $x_1$.  The $x_2$ dependence of the effective free energy in Eq.~\eqref{eq:f-landau-2d-example} is chosen as follows.

First, the function $\gamma(x_2, T)$  determines the minima with respect to $x_2$ in the hot phase $T>T_c$, where $x_1^{\rm eq}=0$.  We set it as
\begin{eqnarray}
\label{eq:gamma}
\gamma(x_2, T) & = & 5 (x_2 - y(T))^2 \nonumber\\
\label{eq:m1-of-T}
y(T) & = & \left( \frac{T}{T_{c}} \right)^2 e^{- \left(T / T_c\right)^2},
\end{eqnarray}
where $T_c$ is the critical temperature of the model \footnote{We set $T_c=1$ in all the figures in this manuscript.}.  For a fixed $T$, the minimum of $\gamma(x_2, T)$ is located at $x_2=y(T)$, which is a non-monotonic function of $T$ with a single maximum at $T_c$, where $ y(T_{c})=x_2^* = e^{-1} \approx 0.37$. 

Next, we construct $\psi(x_2,T)$ such that (i) It generates the second order phase transition at $T=T_c$; (ii) It does not alter the non-monotonic behavior of $x_2^{\rm eq}(T)$; (iii) At low temperatures (below $T_c$) the stability of symmetry line $x_1=0$ changes as a function of $x_2$ in the range of the equilibrium values $(0, x_2^*)$. Namely, the stability point $x_2^{\rm st}$ satisfies $0 < x_2^{\rm st} < x_2^*$.

To have a phase transition in the $x_1$ coordinate at $T_c$, we require that 
\begin{eqnarray}
\label{eq:psi-condition-1}
\psi\Big(x_2^{\rm eq}(T),T)\Big)\begin{cases}
  >0, & \hbox{for } T>T_c \\
  =0, & \hbox{for } T=T_c \\
  <0, & \hbox{for } T<T_c.
\end{cases}
\end{eqnarray} 
In addition, to make the equilibrium of $T^i>T_c$ stable in the $x_1$ direction for dynamics with bath temperature at $T^f<T_c$, we require that $\psi(x_2^{\rm eq}(T^i),T^f)>0$. Combining this condition with the condition in Eq.~\eqref{eq:psi-condition-1} for $T=T^f$, we find these two conditions
\begin{align}
\psi\left(x_2^{\rm eq}(T^i),T^f\right) & >0\nonumber\\
\label{eq:psi-condition-2}
\psi\left(x_2^{\rm eq}(T^f),T^f\right) & <0.
\end{align}
The conditions in Eqs.~(\ref{eq:psi-condition-1}, \ref{eq:psi-condition-2}) are demonstrated graphically in Fig.~\ref{fig:free-energy}(a, c).

\begin{figure}[ht]
  \centering
  \includegraphics[width=1.0\linewidth]{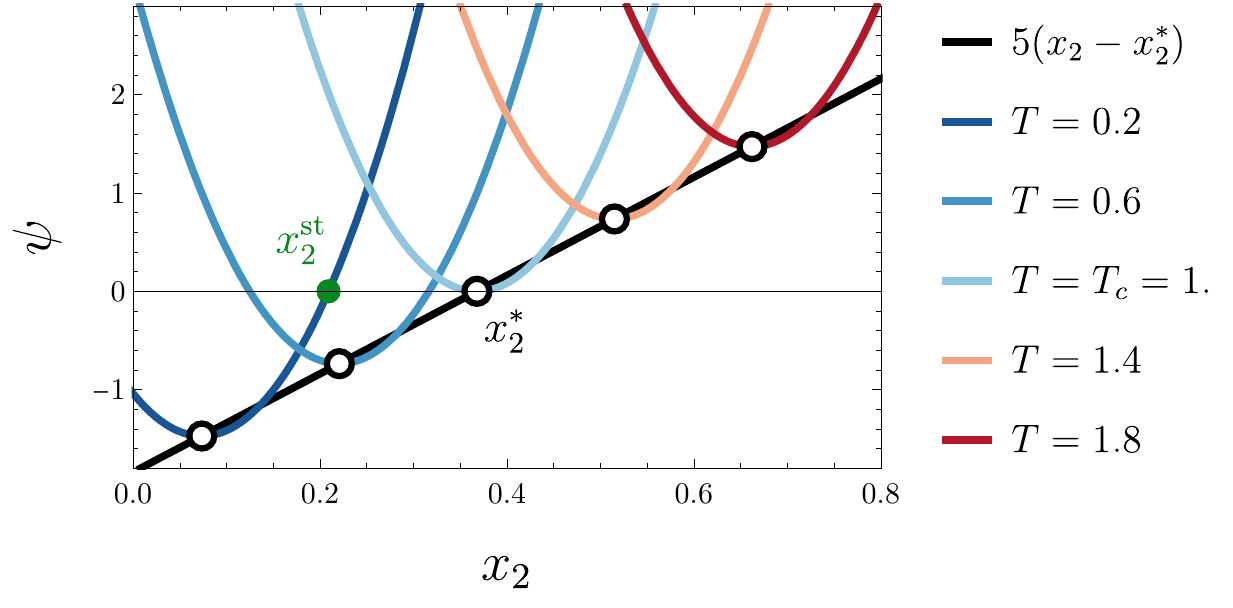}
  \caption{\label{fig:psi2-parabola} $\psi(x_2, T)$, as defined in Eqs. (\ref{eq:psi2}, \ref{eq:a}), is a parabola that slides on the linear line $5(x_2 - x_2^*)$ (black) as a function of temperature.
  The minimum of the parabola coincides with the black line and is denoted by white dots.
  For $T>T_c$, $\psi$ is strictly positive; for $T=T_c$, $\psi$ is non-positive only at $x_2=x_2^*$, where it zeros; for $T<T_c$, $\psi$ is both positive and negative. The contrast between these positive and negative regions at $T<T_c$ exactly accomplishes the wanted behavior in Fig.~\ref{fig:free-energy}(a), which is encapsulated by Eq.~\eqref{eq:f-stability-cond}. The point $x_2$ where $\psi$ vanishes corresponds to the stability point $x_2^{\rm st}$ (such a point is denoted in green in the Fig. for $T=0.2$) that separates between the stable region and the unstable region. It corresponds to the black line in Fig.~\ref{fig:free-energy}(a-b). }
\end{figure}

A simple way to fulfil all the demands for $\psi(x_2, T)$ is by a parabola in $x_2$ that changes as a function of temperature:
\begin{align}
\label{eq:psi2}
\psi(x_2, T) = a x_2^2 + b(T) x_2 + c(T).
\end{align}
The temperature dependence of the parabola is captured graphically by Fig.~\ref{fig:psi2-parabola}.
For $T>T_c$, $\psi(x_2,T)>0$ for all values of $x_2$. At $T=T_c$, it is positive at all values of $x_2$ except for $x_2=x^*_2$, where $\psi(x_2^*,T_c)=0$. For $T<T_c$, it is negative for some values of $x_2$, including $x_2^{\rm eq}(T)$, but positive for some $x_2<x_2^*$.
Note that the roots of $\psi(x_2, T)$, which exist only for $T \le T_c$, determine the boundaries of the stability regions. 
For low enough temperatures, the smaller root of $\psi(x_2, T)$ is negative, and so only the larger root is in the range of the model's parameters.
Thus this single root is exactly the stability point $x_2^{\rm st}$, see for example the green dot in Fig.~\ref{fig:psi2-parabola}.
A concrete choice that adheres to this behavior is given by
\begin{align}
\label{eq:a}
  a & = 80 \nonumber \\
  b(T) & = -10 a x_2^* \frac{T}{T_c} \\
  c(T) & = 5 x_2^* \left( \frac{T}{T_c} - 1 \right) + 5 a (x_2^*)^2 \left( \frac{T}{T_c} \right)^2\nonumber.
\end{align}
We note that the $f$ constructed above is at most quadratic in $x_2$, and that the coefficients of $x_2^2$ in both $\gamma$ and $\psi$ are positive -- therefore the free energy has a minimum at all temperatures.

\subsection{Existence of the Mpemba Effect}

The Landau free energy $f(x_1,x_2;T=0.1)$ given in Eqs. (\ref{eq:f-landau-2d-example}, \ref{eq:gamma}, \ref{eq:psi2}, \ref{eq:a}) is plotted in Fig.~\ref{fig:free-energy}(a). 
By complying with the features explained above, it demonstrates the Mpemba effect: the blue and red dots correspond to $\vec x^{\rm eq}(T^i_{\rm cold})$ and $\vec x^{\rm eq}(T^i_{\rm hot})$, respectively. 
In both cases, following the dynamic in Eq.~\eqref{eq:langevin}, $x_2$ decreases as a function of time, and reaches the stability line (the black line in Fig.~\ref{fig:free-energy}(a)) at finite time. Crossing from the stable region to the unstable region, $x_1$ is deflected from the $x_1=0$ line. 
As the cold (blue) initial condition  $\vec x^{\rm eq}(T^i_{\rm cold})$ passes by the hot (red) initial condition  $\vec x^{\rm eq}(T^i_{\rm hot})$, and they follow the same dynamic, it takes longer time for the cold initial condition to reach the phase transition line than the hot initial condition.
Thus the Mpemba effect exists.

Finally, let us analyze the range of temperatures for which the Mpemba effect occurs. The phase transition time is a function of both the final  and initial temperatures, namely $t_c(T^i, T^f)$. If for some value of $T^f$, the phase transition time has $t_c(T^i_{\rm hot}) < t_c(T^i_{\rm cold})$ (which means that the hot system crosses to the ordered phase faster), then we have a Mpemba effect. Recall that in this analysis we require that $T^i>T_c$ and $T^f<T_c$.

To understand the dependence of the phase transition time $t_c$ on the initial temperature $T^i$ and the bath temperature $T^f$, note that $T^i$ sets the initial value of $x_2^0\equiv x_2(t=0)$, and $T^f$ sets the stability line $x_2^{\rm st}$ (recall Eq.~\eqref{eq:f-stability-cond}).
As $x_2^{\rm eq}(T^i)$ is a decreasing function (for $T^i>T_c$), increasing $T^i$ means that the initial value $x_2^0$ decreases and starts closer to $x_2^{\rm st}$. 
Thus we expect that for high enough $T^i$, the initial condition starts in the unstable region, $x_2^0 < x_2^{\rm st}$, and so the phase transition happens instantaneously, namely $t_c=0$.

\begin{figure}[ht]
  \centering
  \includegraphics[width=1.0\linewidth]{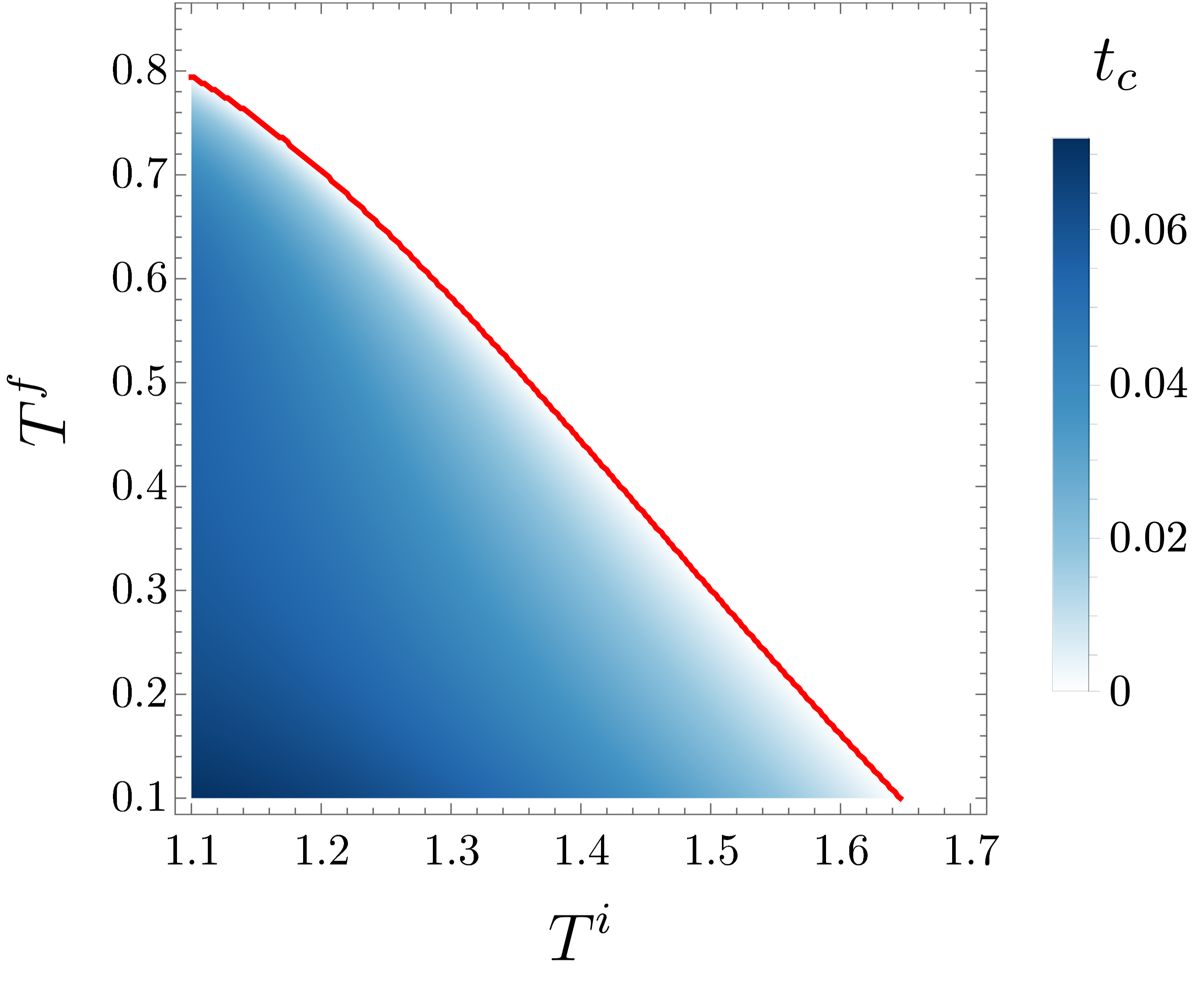}
  \caption{\label{fig:tc} The phase transition time $t_c$ as a function of the initial temperature $T^i$ and the final temperature $T^f$. Note that the initial conditions correspond to the disordered phase so $T^i>T_c$, and the final temperatures correspond to the ordered phase so $T^f < T_c$. As for any final temperature $T^f \lessapprox 0.8$, the phase transition time  $t_c(T^i, T^f)$ is a decreasing function of $T^i$, the system exhibits the Mpemba effect. The red line separates the temperatures space $(T^i, T^f)$ to two regions: on the left the Mpemba effect exists, whereas on the right it does not.  }
\end{figure}

Next, consider the dependence of $t_c$ on the bath temperature $T^f$.
As $T^f$ increases, the position of the stability line $x_2^{\rm st}$ increases as well.
To see this, note that the stability line $x_2^{\rm st}$ is given by the greater root of $\psi(x_2, T^f)$ (Eq.~\eqref{eq:f-landau-2d-example}).
Increasing $T^f$ increases the roots of $\psi(x_2, T^f)$, as can be seen in Fig.~\ref{fig:psi2-parabola}.
Therefore, increasing $T^f$ means that the range of initial temperatures which have non-zero $t_c$ gets smaller.
These two features are apparent in Fig.~\ref{fig:tc}, where $t_c$ is plotted as a function of $T^i$ and $T^f$. The Mpemba effect therefore exists for all triplets $T^f<T^i_{\rm cold}<T^i_{\rm hot}$ such that the point $(T^f,T^i_{\rm cold})$ is to the left of the red line in Fig~.\ref{fig:tc}.

\section{Discussion}
\label{Sec:discussion}

In this manuscript, we used the gradient of the free energy as the force that drives the macroscopic parameters $\vec x(t)$ in the thermal relaxation process, and identified the point in time at which the stability of order parameter changes as the ``phase transition time". With this identification we could define a Mpemba effect with respect to a phase transition, and construct a Landau free energy that demonstrates it.  We stress that the Mpemba effect defined here is not restricted to the specific dynamics we considered (Eq.~\eqref{eq:langevin}). Indeed, as stability of the order parameter is generic in second order phases transitions, the identification of the time in which the order parameter crosses from a stable region to an unstable region can be applied in other non-equilibrium relaxation models as well. We demonstrate this in the Supp. Info., where we consider the phase transition time for a model with a microscopic dynamics whose thermodynamic limit is not the gradient of the free energy.

Of specific interest are non mean-field models, which correspond to more realistic systems with spatial structure. In these models a spatially dependent field is needed for a proper description. Analyzing the dynamic in Fourier components, which are commonly coupled, it might happen that the non-zero components corresponding to spatial fluctuations of the field affect the zero Fourier component, which determines the mean of the order parameter field. If the stability of the zero Fourier component changes due to the dynamics of other Fourier components, a phase transition at a finite time, and consequently -- a Mpemba effect might exist even in a system with a single macroscopic parameter as the ferromagnetic Ising model \cite{Vadakkayil2021}. In other words, the non-zero Fourier components in statistical field theories can play the same role as $x_2$ plays in the simple example discussed in this manuscript. 

The phase transition time defined in Eq.~\eqref{Eq:PhaseTRansTime} is not the only possible definition. For example, an alternative definition that can be used is the point in time at which the probability distribution of the order parameter changes from having a single maximum at $x_1=0$ to having two distinguishable maxima at non-zero value of the order parameter. This phase transition time is expected to be correlated, but delayed with respect to the phase transition time used in this manuscript. The main advantage in such a definition is that it is experimentally and numerically easier to observe in models where direct stability analysis cannot be done. However, in this case the exact phase transition time depends on the noise characteristics.

In contrast to the Mpemba effect in Markovian systems \cite{Lu2017,Kumar2020} or in granular gases \cite{Lasanta2017,torrente2019large}, the inverse Mpemba effect -- where a cold system heats faster than a hot one -- is less expected in the suggested framework. In the regular effect, the two hot systems are initiated at $x_1^{\rm eq} = 0$ and the phase transition happens at $t_c$ where $ x_1(t_c)=0$ becomes unstable. In the inverse effect, the two systems are expected to be initiated at some $x_1^{\rm eq}\neq 0$. Regardless of the stability in the $x_1$ direction, most models cannot attain $\langle x_1(t) \rangle=0$ at a finite time $t$, but only approach zero asymptotically at $t\to\infty$. Therefore it is not obvious how to identify the exact phase transition time in this case. The inverse Mpemba effect might nevertheless exist in this framework, but in models that have a second order phase transition between a disordered phase  at cold temperature and an ordered phase at high temperature. An example for such a model is the mean-field anti-ferromagnet at some small range of magnetic field values \cite{vivesUnifiedMeanfieldStudy1997}.  

The model presented in this manuscript is phenomenological, and it would be of great interest to find a concrete, microscopic model that demonstrates the same effect. However, the temperature dependence in the free energy we constructed is quite involved. It cannot originate from a simple coarse-graining procedure that gives a linear temperature dependence as e.g. in   \cite{vivesUnifiedMeanfieldStudy1997}, but rather from a more involved procedure, e.g. the Hubbard-Stratonovich transformation \cite{hubbard1959calculation}, that often results in a more complicated temperature dependent free energy.

Lastly, we note that our discussion here is limited to a second order phase transition, whereas in various examples as water \cite{Jeng2006} or clathrate hydrates \cite{paper:hydrates} the observed Mpemba effect happens through a first order phase transition. The non-equilibrium dynamic through a first order phase transition, e.g. ``nucleation and growth'' \cite{gillespie1981stochastic,PhysRevE.56.5615}, is vastly different from the dynamic discussed here, and is a main challenge for future studies.

\section{Acknowledgements}
We would like to thank David Mukamel, Gianluca Teza, Shahaf Aharony, Samuel Safran and Hillel Aharoni for useful discussions. O. R. is the incumbent of the Shlomo and Michla Tomarin career development chair, and is supported by the Abramson Family Center for Young Scientists, the Israel Science Foundation Grant No. 950/19 and by the Minerva foundation.

\bibliography{mpemba_phase_transitions}

\appendix
\section{Example for a Phase Transition at Finite Time in a Mean Field Model}
\label{sec:appendix-phase-trans-finite-time}

The phase transition time $t_c$ defined in Eq.~\eqref{Eq:PhaseTRansTime} is quite intuitive, but to the best of our knowledge, it is not commonly used. In this section we explore its nature, and show that it behaves as one expects from a phase transition time. To this end, we first note that $t_c$ defined in Eq.~\eqref{Eq:PhaseTRansTime} is already an average quantity. To understand the validity of the definition, we therefore define the (stochastic) ``empirical phase transition time" for a given realization as the minimal $\tilde t_c$ that solves
\begin{eqnarray}
\partial_{x_1}^2f\Big((x_1=0,x_2(t));T_b\Big)=0
\end{eqnarray}

The suggested definition in Eq.~\eqref{Eq:PhaseTRansTime} is physically solid only if the mean time, $\langle \tilde{t}_c \rangle$, does not diverge and its variance decreases with the system size. However, these cannot be checked at the level of a Landau theory, where the noise is somewhat synthetically added, and the thermodynamic limit is already taken. Instead, it should be considered at the microscopic level. As we demonstrate in what follows, crossing the phase transition at a finite time can be demonstrated for example in the Glauber dynamics of the mean field anti-ferromagnetic Ising model, discussed e.g. in \cite{vivesUnifiedMeanfieldStudy1997}. This example also serves us in demonstrating that our definition for the phase transition time works not only for the gradient of the free energy dynamics (Eq.~\eqref{eq:langevin}), but also for other possible dynamics that can arise from microscopic models.

\subsection{Model Definition}
\label{sec:definition-model}

To demonstrate the well behavior of the phase transition time and the applicability of our definition in non free energy gradient decent dynamics, we consider the mean field model of the Ising anti-ferromagnet under Glauber dynamics. The equilibrium properties of the model are presented in \cite{vivesUnifiedMeanfieldStudy1997}, and the Glauber dynamic for this model is discussed in \cite{Klich2019}.

In mean field models of spin systems, every spin interacts with all other spins in the system.
As we consider the anti-ferromagnet, we divide the system into two sub-lattices of equal size where every spin in one sub-lattice interacts with all other spins in the other sub-lattice, but not with the spins on the same sub-lattice.
In the mean field picture there is no spatial structure, thus the state of the system can be described by the number of up-spins in the first sub-lattice $N_{1, \uparrow}$ and the number of up-spins in the second sub-lattice $N_{2, \uparrow}$.
For a system of $N$ spins, each sub-lattice is composed of $N / 2$ spins, and so the normalized magnetization of each sub-lattice is given by
\begin{equation}
\label{eq:3}
y_1 = \frac{N_{1, \uparrow} - N_{1, \downarrow} }{N / 2}, \qquad y_2 = \frac{N_{2, \uparrow} - N_{2, \downarrow} }{N / 2},
\end{equation}
where $N_{i, \uparrow}, N_{i, \downarrow}$ are the number of up-spins and down-spins in sub-lattice $i$, respectively.
In terms of $y_1, y_2$ the mean field Hamiltonian is given by 
\begin{equation}
\label{eq:6}
H = - N \left( J y_1 y_2 + h (y_1 + y_2) \right),
\end{equation}
where $J$ is the coupling constant and $h$ is the magnetic field.
For anti-ferromagnetic interactions, the coupling constant is negative, and we set $J=-1$ for simplicity.
From these magnetizations of the sub-lattices $y_1, y_2$, follow the more informative parameters of staggered magnetization $s$ and total magnetization $m$ defined as
\begin{equation}
  \label{eq:5}
 s = \frac{y_1 - y_2}{2}, \qquad m = \frac{y_1+y_2}{2}.
\end{equation}
As the order of the system is encapsulated by the staggered magnetization only, using our notation of Sec. \ref{sec:2d-phase-transition}, the order parameter is $x_1=s$, and the other macroscopic parameter is $x_2=m$.
Indeed, for $h<1$ the staggered magnetization $s$ satisfies Eq.~\eqref{eq:x1-order-disorder}, namely 
\begin{equation}
\label{eq:s-order-disorder}
s^{\rm eq}(T)
\begin{cases}
  = 0, \quad \text{disordered, } & T\geq T_c \\
  \neq 0, \quad \text{ordered, } & T< T_c.
\end{cases}
\end{equation}
The critical temperature $T_c$ is a function of the magnetic field $h$, and it exists for small enough values of $h$. In what follows we assume that $h$ is small enough for $T_c$ to exist.

\subsection{Glauber Dynamics of the Model}
\label{sec:glaub-dynam-model}

The Glauber dynamics for this system, allowing only single spin flips, was derived in Ref.  \cite{Klich2019}.

\begin{figure}[ht]
  \centering
   \includegraphics[width=1.0\linewidth]{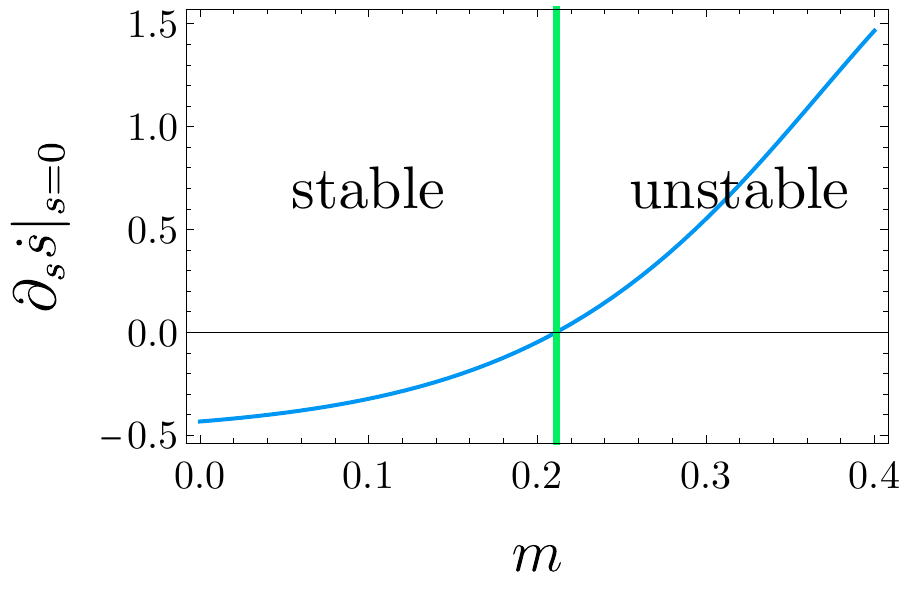}
  \caption{\label{fig:antiferro-sdot-stability} The stability of the flow on the symmetry line $s=0$, $\frac{\partial \dot{s} }{\partial s} \Big|_{s=0, m}$ given in Eq.~\eqref{eq:sdot-stability-cond} for $h=0.5, T_b=0.2$. The green vertical line separates between the stable and the unstable regions. }
\end{figure}

In the thermodynamic limit, the dynamical equations for $s, m$ are given by
\begin{align}
\label{eq:sdot}
\dot{s} &= \frac{1}{4} \left[ \tanh \left( \frac{h-m+s}{T_{b}} \right) - \tanh \left(\frac{h-m-s}{T_{b}} \right) \right] -\frac{s}{2} \\
  \label{eq:mdot}
\dot{m} &= \frac{1}{4} \left[ \tanh \left( \frac{h-m-s}{T_{b}} \right) - \tanh \left(\frac{h-m+s}{T_{b}} \right) \right] -\frac{m}{2} ,
\end{align}
where $T_{b}$ is the bath temperature.
This dynamic is not the gradient flow of any potential, as can be easily checked. Therefore, it provides a different type of non-equilibrium relaxation dynamic than considered in the main text (Eq.~\eqref{eq:langevin}). Nevertheless, an analogous stability criterion to the one in Eq.~\eqref{Eq:PhaseTRansTime} can be defined in this system too.
Indeed, the stability of $s$ at $s=0$ is determined by the derivative of $\dot{s}$ in the $s$ direction: $\partial_s \dot{s} |_{s=0, m} < 0$ corresponds to $m$ values which are stable with respect to $s$, whereas $\partial_s \dot{s} |_{s=0, m} > 0$ corresponds to $m$ values which are unstable with respect to $s$.
The explicit condition, using Eq.~\eqref{eq:sdot}, is given by
\begin{equation}
\label{eq:sdot-stability-cond}
\frac{\partial \dot{s} }{\partial s} \Big|_{s=0, m} = \frac{1}{2} \left( \frac{1}{T_b \cosh^2 \left( \frac{h-m}{T_{b}} \right)} - 1 \right) \begin{cases}
  < 0, \qquad \text{stable} \\
  > 0, \qquad \text{unstable}.
\end{cases}
\end{equation}
This function is plotted in Fig.~\ref{fig:antiferro-sdot-stability} for $T_b=0.2$ and $h=0.5$, and it shows the two stability regions.
To see how this stability condition corresponds to a measurement of the ``phase transition time'' $t_c$ we next consider the equilibrium line in the configuration space of the macroscopic parameters $s, m$.

\subsection{The Equilibrium Line}
\label{sec:equilibrium-line}

The equilibrium line of the model is obtained by finding the fixed points of the dynamics in Eqs.~(\ref{eq:sdot}, \ref{eq:mdot}), i.e. by solving for $\dot{s} = 0, \dot{m} = 0$ (for an alternative method see \cite{vivesUnifiedMeanfieldStudy1997}).
The equilibrium line and the dynamic properties of the model are a function of the magnetic field $h$.
For weak magnetic fields, $|h| < |J|$, these properties are qualitatively the same, and therefore from now on we set $h=0.5$ for all the numerical calculations that are presented.

\begin{figure*}[ht]
  \centering
  \includegraphics[width=1.0\linewidth]{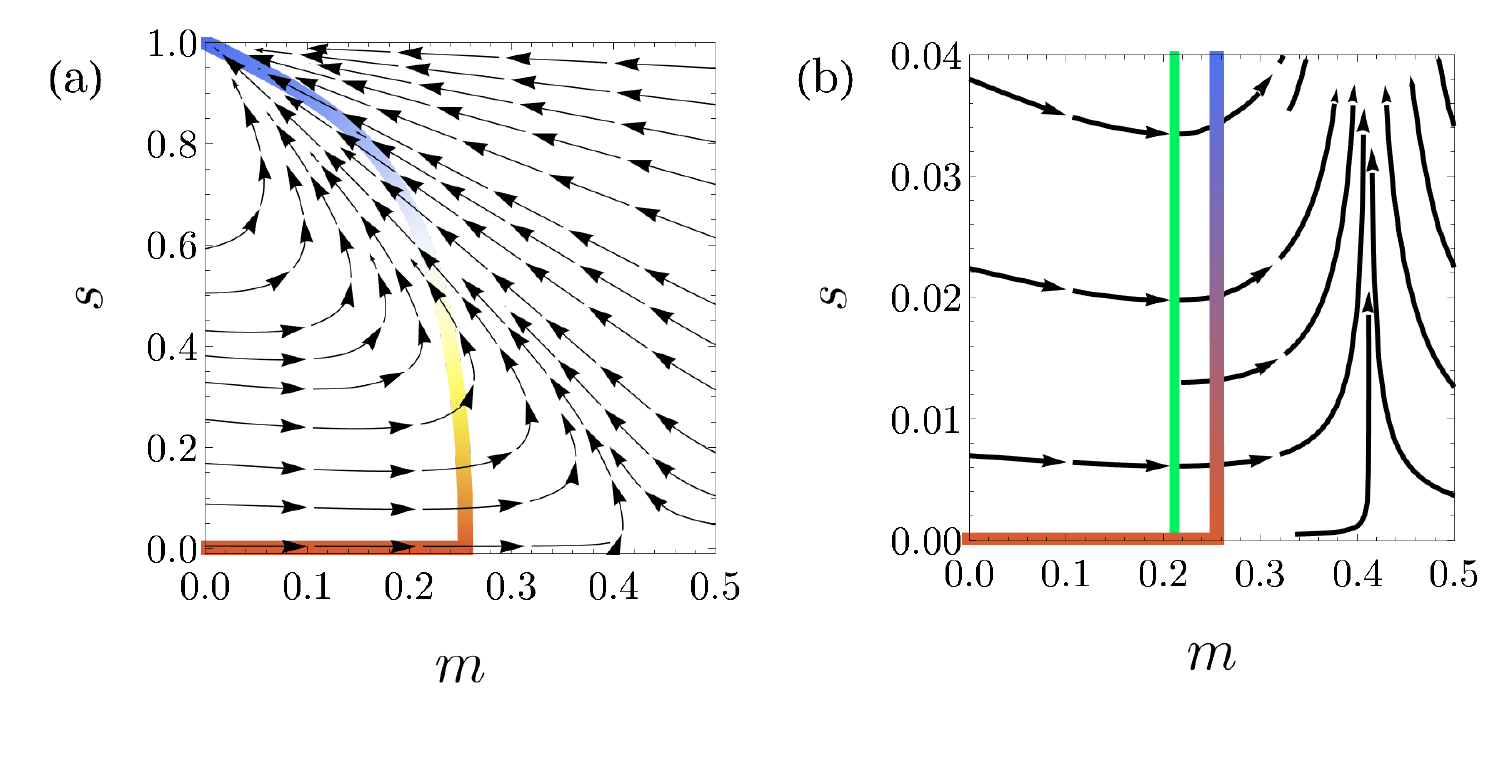}
  \caption{\label{fig:antiferro-equilibrium}
  (a) The equilibrium line of the anti-ferromagnet for $h=0.5$. The equilibrium line corresponds to solving $\dot{s} =0, \dot{m} =0$ given in Eqs.~(\ref{eq:sdot}, \ref{eq:mdot}). Temperatures range from $T=\infty$ to $T=0$ that correspond to red and blue, respectively. The arrows indicate the flow of the macroscopic parameters for the bath temperature $T_b=0.2$.
  (b) Inset of panel (a) showing the stability of the flow $\dot{s}, \dot{m}$ with respect to the symmetry line $s=0$ at bath temperature $T_b=0.5$. The red-blue line is the relevant part of the equilibrium line which is shown fully in panel (a). The green line denotes the separation of the two stability regions: the left part is the stable region for which $\partial_s \dot{s} |_{s=0, m} < 0$, and the right part is the unstable region for which $\partial_s \dot{s} |_{s=0, m} > 0$, see Eq.~\eqref{eq:sdot-stability-cond} and Fig.~\ref{fig:antiferro-sdot-stability}. The flow, denoted by the black arrows, shows the stability trends. All initial conditions on the left of the green line have finite non-zero ``phase transition time'' $t_c$.}
\end{figure*}

The equilibrium line, plotted in Fig.~\ref{fig:antiferro-equilibrium}(a), has the same qualitative shape as the example given in Sec.~\ref{sec:example-mpemba2-2d}.
In particular the non-monotonicity of the equilibrium values of the magnetization as a function of temperature, $m^{\rm eq}(T)$, which has a maximum at $T_c$ is qualitatively the same.

\subsection{Stability of the Non-Equilibrium State When Quenched to a Cold Temperature}
\label{sec:stab-non-equil}

To show that the phase transition time $t_c$ is well defined, we perform Monte Carlo simulations on finite systems with different sizes, from which we measure numerically the statistics of $\tilde t_c$.

The numerical measurements are performed as follows.
For each system size $N$, we initiated each of the $3\times 10^4$ realizations with a random spin configuration. This is equivalent to sampling the system from the equilibrium associated with $T^i= \infty$. Each realization is then evolved by a Monte-Carlo algorithm that implements the Glauber dynamics with $T_b=0.2$. Note that the model is stochastic by its discrete nature, so no added noise is needed. The initial condition corresponds to $(s^{\rm eq}(T=\infty), m^{\rm eq}(T=\infty)) = (0, 0)$, which is a stable point in the $s$ coordinate.
By Eq.~\eqref{eq:sdot-stability-cond}, for $h=0.5, T_b=0.2$, we find that the value of $m$ which separates the stable and unstable regions is given by $m^{\rm st}(T_b=0.2) \approx 0.211$ (see Fig.~\ref{fig:antiferro-sdot-stability}).
For each realization, we track the evolution of $(s(t), m(t))$ and we denote the time in which the system crosses $m^{\rm st}$ for the first time as the ``phase transition time'' $\tilde t_c$.
To compare between different system sizes we count $\tilde t_c$ in units of ``Monte-Carlo Sweep time'', namely the number of spins $N$. 

\begin{figure*}[htbp]
  \centering
  \includegraphics[width=0.9\linewidth]{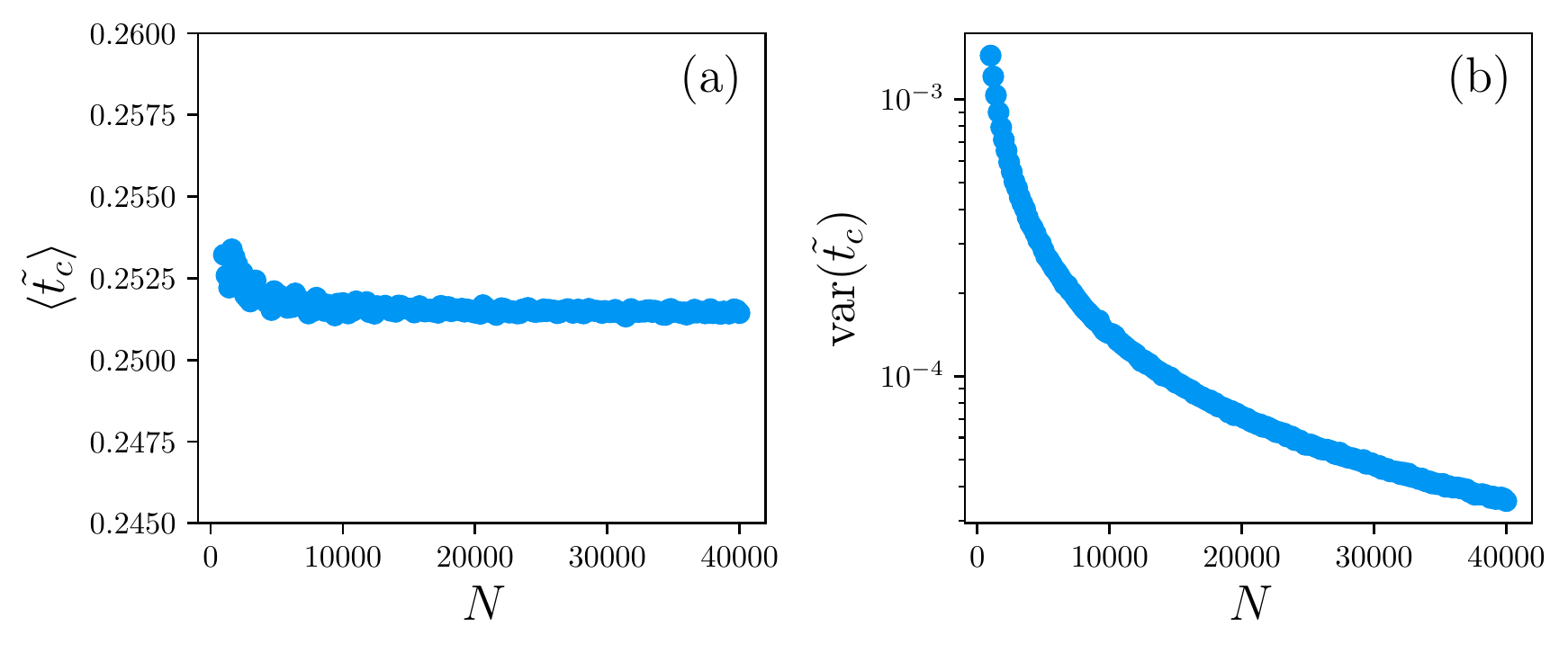}
  \caption{The mean (a) and the variance (b) of the ``phase transition time'' $\tilde t_c$ as a function of the system size $N$. The time $\tilde t_c$ is measured in units of Monte-Carlo sweeps, i.e. in units of $N$. The mean is approximately constant, and the variance decreases as $N$ increases. Thus we conclude that $\tilde t_c$ is a well behaving quantity in the thermodynamic limit.}
  \label{fig:tc-antiferro}
\end{figure*}

Figure~\ref{fig:tc-antiferro} shows the calculated mean and variance of $\tilde t_c$. It can be seen that the mean is constant, whereas the variance decreases with the system size.
Thus we conclude that the stochastic variable $\tilde t_c$ is a measurable quantity which gives a finite non-zero time for the crossing of the phase transition.

\end{document}